\documentstyle[11pt,paspconf,epsf]{article}

\begin{document}

\def\funit{erg~cm$^{-2}$~s$^{-1}$~$\AA^{-1}$}

\title{HUT Observations of Time-Variable Absorption in NGC 4151}

\author{ }
\affil{Department of Physics and Astronomy, and Center for Astrophysical
Sciences, Johns Hopkins University, Baltimore MD 21218}

\begin{abstract}
We present the results of ultraviolet spectral monitoring of the Seyfert
galaxy NGC 4151 performed by the Hopkins Ultraviolet Telescope during
the Astro-2 mission.  We show how the response of the Lyman continuum
optical depth to variations in the continuum flux may be used to derive
estimates of the distance of the absorbing gas from the central object.
In this case, it seems that the absorbing gas is located $\leq 50$~pc
away from the central source.
\end{abstract}


\section{Introduction}

    Most work analyzing the absorption lines that tell us about AGN outflows
is done on single observations.  We then study the detailed shapes of the
line profiles to infer the character of the gas as a function
of line-of-sight velocity.  As is evident by a comparison of the different
models discussed at this meeting, it is very difficult to attach a radial
scale to these inferences.

    Time-variability offers a valuable new perspective, particularly
when the ionization state of the absorbing gas is controlled by photoionization
due to the AGN continuum.  Unlike the case of variability in emission lines,
there is no ambiguity about whether the photoionized gas sees a different
continuum than the one we do---the continuum directed at the gas is
the same continuum shining at us, and there is identically zero
time delay between the luminosity we see now and the luminosity incident
upon the gas when it did the absorbing.  For these reasons, the interpretation
of time-variable absorption lines is far more direct than the interpretation
of time-variable emission lines.

\section{Observations}

\begin{figure}[h]
\plotfiddle{"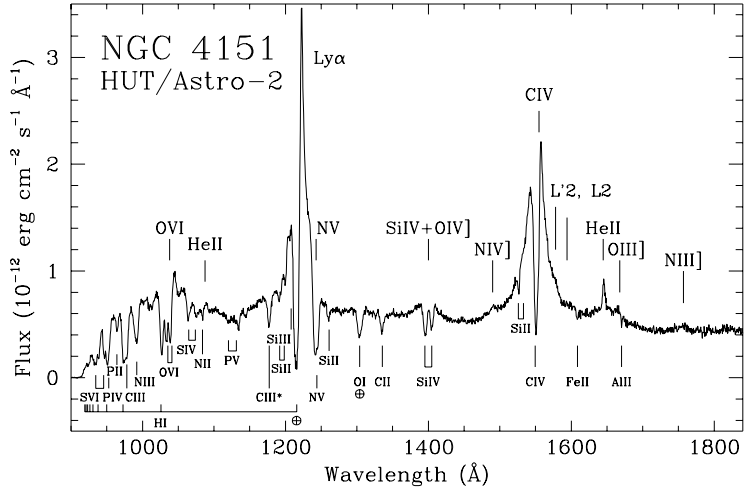"}{3.0in}{-90}{40}{40}{-175}{230}
\caption{A global mean spectrum of NGC 4151 based on all HUT-2 observations
of this object.  The principal lines are identified.  Those labelled
$\oplus$ are geocoronal.} \label{fig-1}
\end{figure}

    To pursue this program (and for several other reasons), we observed
the bright type 1 Seyfert galaxy NGC 4151 with the Hopkins Ultraviolet
Telescope (HUT) six times over a span of 11 days during the Astro-2 mission
of February/March, 1995.  To show the quality of the data we obtained, we
first present a spectrum based on a sum of all our data (Figure~\ref{fig-1},
from Kriss et al. 1995).
As can be easily seen, the signal/noise ratio of these data are high enough
to permit the identification of large numbers of lines, both in absorption
and in emission, including some that are quite weak.  The absorption
lines seen span a wide range in ionization, from O$^{+5}$ and N$^{+4}$
to C$^{+1}$, Si$^{+1}$, and Fe$^{+1}$.  What is not so easily
seen in this figure is that there is definite Lyman continuum
absorption at a redshift of $320 \pm 60$~km~s$^{-1}$; that is, blueshifted
by $\simeq 700$~km~s$^{-1}$) from the systemic redshift of NGC 4151.

   The luminosity of NGC 4151 varied substantially over the course of
these observations.  If we take as a fiducial continuum level the flux
at 1450\AA, we saw the flux rise from $\simeq 4.5 \times 10^{-13}$\funit
on 4 March to almost double that ($\simeq 8 \times 10^{-13}$\funit) on
7 March, and then watched it drop back down to a level only slightly higher
than at the beginning of this experiment.

\begin{figure}[h]
\plotfiddle{"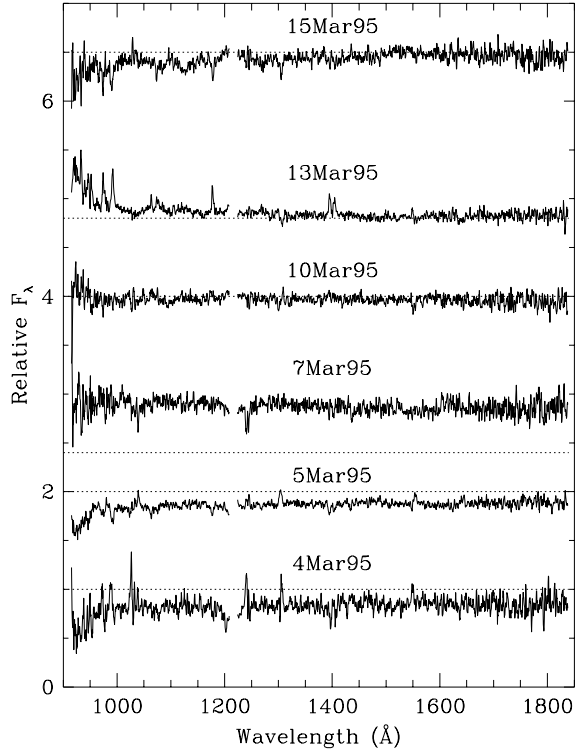"}{3.0in}{0}{40}{40}{-150}{-10}
\caption{Each separate spectrum, normalized to the mean.  Unit ratio is shown
by the dotted line near each spectrum.} \label{fig-2}
\end{figure}

   A particularly noteworthy feature of the spectral time-variability was
the change in Lyman edge optical depth.  This is best seen in a sequence
of normalized spectra (Figure~\ref{fig-2}), in which each day's spectrum
is displayed as a ratio to the mean spectrum.
While the shape of the continuum and the equivalent widths of both the emission
and the absorption lines changed very little, the relative flux in the Lyman 
continuum clearly swings sharply up and down.

\begin{figure}[h]
\plotfiddle{"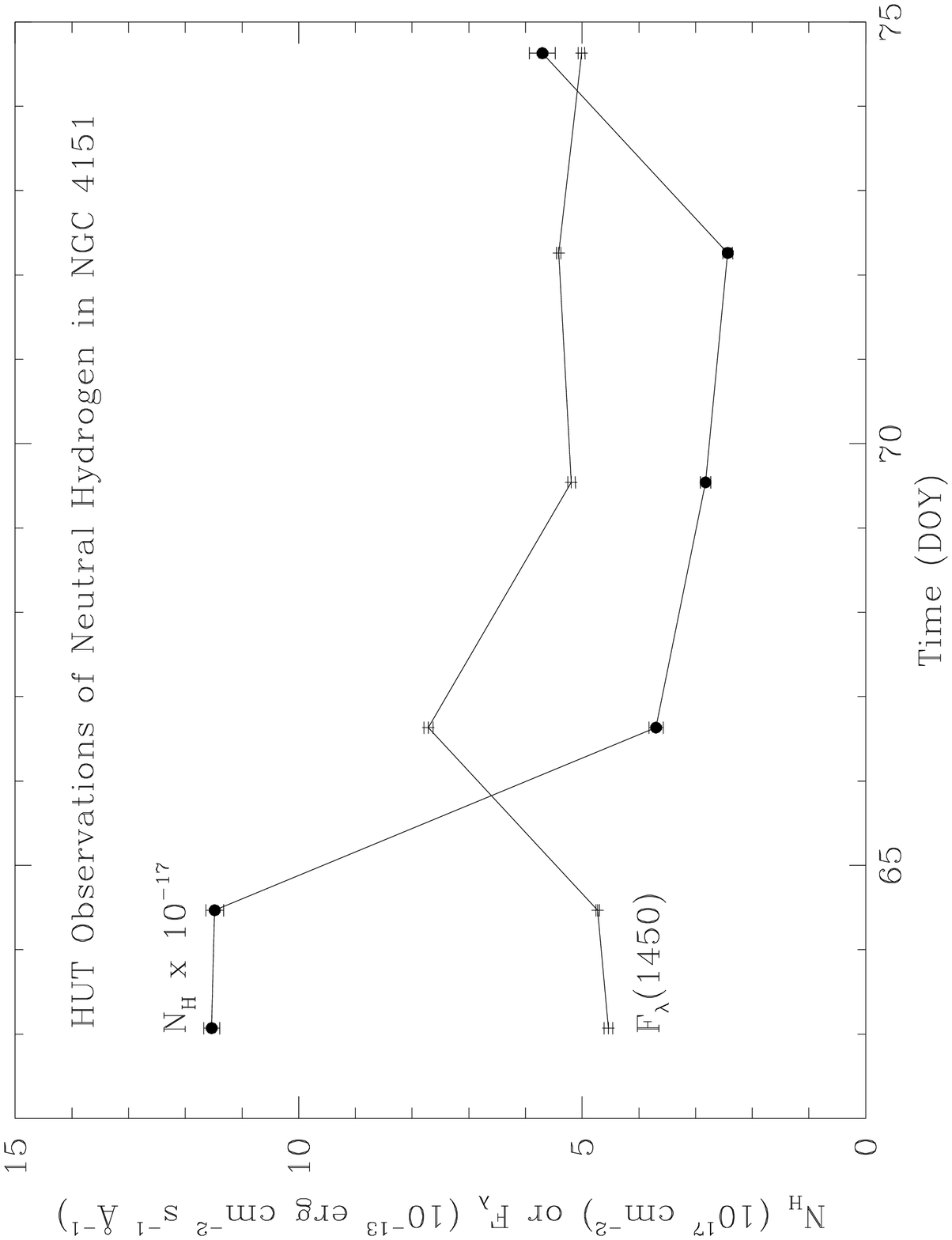"}{3.0in}{-90}{40}{40}{-160}{250}
\caption{The variation in the measured neutral hydrogen column
density from HUT observations of NGC 4151 during Astro-2 is shown along with
the measured continuum intensity at 1450 \AA.
The x-axis shows time labelled by day of the year 1995.} \label{fig-3}
\end{figure}

    The drop in flux at the redshifted Lyman edge is easily translatable into a
column density of neutral hydrogen atoms.  The variations in this column
density {\it vis-a-vis} the variations in the continuum flux are shown in
Figure~\ref{fig-3}.
The sharp rise in the continuum near the beginning of the observations
is mirrored by an equally sharp fall in the HI column density, while the
slow decline in the continuum power towards the end of the observations is
reflected in a slow, and delayed, recovery in the HI column.  At least at
the qualitative level, such behavior is exactly what one would expect from
gas whose neutral fraction is regulated by exposure to the central engine's
ionizing continuum.

\section{Ionization Analysis}

   We can investigate this relationshiop more quantitatively by examining
the detailed relationship between the time-variation of the neutral H
density and the ionizing continuum:
\begin{equation}
{d n_{HI} \over dt} = n_e n_p \alpha_{rec} - n_{HI} \int_{\nu_H} \, d\nu  \,
{L_\nu \over 4\pi r^2 h\nu} \sigma_{ion}(\nu),
\end{equation}
where $n_{e,p}$ are the electron and proton densities, $\alpha_{rec}$ is
the radiative recombination coefficient, $\nu_H$ is the frequency of the
Lyman edge, $L_\nu$ is the luminosity per unit frequency in the continuum,
$r$ is the distance of the absorbing gas from the source of radiation, and
$\sigma_{ion}$ is the photoelectric cross section.

     It is convenient to
divide through equation 1 by $n_{HI}$, to yield
\begin{equation}
{d\ln n_{HI} \over dt} = n_e \alpha_{rec} \Phi^{-1} - {F_{ion} \over
h\nu_H} \langle \sigma_{ion}\rangle,
\end{equation}
where $F_{ion}$ is the integrated ionizing flux at the location of the gas,
$\langle \sigma_{ion}\rangle$ is the photoelectric cross section
appropriately averaged over the ionizing spectrum, and $\Phi = n_{HI}/n_p$.
When the equation for the rate of change of the neutral density is written
in this form, it is apparent
that the timescale for change of the column density $t_*$ is simply
\begin{equation}
t_{*}^{-1} = t_{rec}^{-1} - t_{ion}^{-1},
\end{equation}
where the recombination timescale is
\begin{equation}
t_{rec} = {\Phi \over n_e \alpha_{rec}}
\end{equation}
and the ionization timescale is
\begin{equation}
t_{ion} = {h\nu_H \over F_{ion}\langle \sigma_{ion}\rangle}.
\end{equation}

    In equilibrium, the ionization and recombination timescales are equal.
When the gas is out of equilibrium in the sense that ionization
dominates recombination (that is, the gas is less ionized than it would
be at the current flux level), $t_* \geq t_{ion}$.  In the limit that
ionization completely dominates recombination, $t_* \rightarrow t_{ion}$.
Since $t_{*}$ depends only on the ionizing flux at the location of the
absorbing gas, and we measure the ionizing flux at Earth, in this situation
$t_*$ may be used to estimate the distance of the absorbing gas from
the central source:
\begin{equation}
r \leq \left({f_{ion} \langle \sigma_{ion}\rangle t_* \over h\nu_H}\right)^{1/2}
D,
\end{equation}
where $f_{ion}$ is the ionizing flux measured at Earth, and
$D$ is the distance from the continuum source to Earth.

     Inference of conditions in the absorber by using $t_*$ during periods
of net recombination is a bit more indirect.  As equation 4 shows, when
$t_* \geq t_{rec}$, we are {\it not} (as is commonly supposed) given
a direct estimate of $n_e$.  Instead, we obtain an estimate of the ratio
$n_e/\Phi$, where the ratio of neutrals to protons $\Phi$ is {\it a priori}
unknown, and could be very different from unity.  However, it is possible
to derive a {\it different} quantity, the departure of the ionization
balance from equilibrium, when periods of both net recombination and
net ionization are observed in the same object.

    The argument begins with the instantaneous equilibrium neutral/ionized
ratio
\begin{equation}
\Phi_{eq}(t) = {n_e \alpha_{rec} h\nu_H \over F_{ion}(t) \langle \sigma_{ion}
\rangle}.
\end{equation}
We assume that $n_e$ is constant on the timescale of the observed
fluctuations.  With this assumption, we may use equation 7 to write $n_e$ in
terms of $\Phi_{eq}(t)$ and $F_{ion}(t)$.  Substituting this expression for
$n_e$ in the definition of $t_{rec}$ gives
\begin{equation}
t_{rec}(t) = {\Phi(t) \over \Phi_{eq}(t)}{h\nu_H \over F_{ion}(t)
\langle\sigma_{ion}\rangle}.
\end{equation}
Since we may estimate $t_{rec}$ from $t_*$ (in the same limiting sense as
when we estimate $t_{ion}$ from $t_*$ during periods of net ionization),
we find
\begin{equation}
{\Phi(t) \over \Phi_{eq}(t)} \leq \left({f_{ion}(t) t_*(t)
 \langle \sigma_{ion}\rangle \over h\nu_H}\right)\left({D^2 \over r^2}\right).
\end{equation}

   Now we apply these arguments to the changing neutral column density and
ionizing flux we observed in NGC 4151.  When the neutral column was falling,
the timescale for a decrease by a factor of $e$ was $\simeq 2$~d or less,
and the observed ionizing flux density at the Lyman edge (after
correction for Galactic extinction) was $\simeq 9 \times 10^{-13}$\funit.
The distance from the ionizing source to the absorbing gas is then
\begin{equation}
r \leq 50\left({D \over 15 \hbox{Mpc}} \right)\hbox{pc}.
\end{equation}
Note that this is an upper bound in two senses:
in the sense we have already discussed, that recombination might partially
cancel ionization; and, because we do not have arbitrarily fine time resolution,
the real $t_*$ might be less than our estimate of $\simeq 2$~d.  On the other
hand, when the neutral column
rose, the $e$-folding timescale was more like 3~d, and the Lyman edge
flux density was slightly smaller, $\simeq 7 \times 10^{-13}$\funit.  Using the
upper bound on $r$ just found, the departure from equilibrium is
\begin{equation}
{\Phi \over \Phi_{eq}} \leq 1.
\end{equation}
On the one hand, this appears to be a tautological result---when
the gas is recombining
$\Phi$ {\it must} be less than $\Phi_{eq}$.  On the other hand, since a
significantly smaller $r$ would lead to a {\it larger} value
of $\Phi/\Phi_{eq}$, we argue that the requirement of consistency suggests
that 50~pc is not a tremendous overestimate.  If so, the gas is also not
too far from equilibrium, even during these column density fluctuations.

\acknowledgments

    This work was partially supported by NASA Contract NAS 5-27000, and NASA
LTSA Grants NAGW-3156 and NAGW-4443.

\begin{question}{Roger Blandford}
Do you think that you can account for the variable absorption by moving the
source---{\it e.g.}, if it is a hot spot on an accretion disk?
\end{question}
\begin{answer}{Julian Krolik}
I think that's unlikely.  Let's make the approximation that the flux we
see is dominated by the moving hot spot.  Then the change in column density
might be attributed to inhomogeneity in the absorber on the scale of the
displacement in the hot spot.  NGC 4151 has a bolometric luminosity
$\sim 10^{44}$~erg~s$^{-1}$.  If it is accreting at 0.1 Eddington, its
central black hole is $\sim 10^7 M_{\odot}$, so the brightest part of the
accretion disk, at $\sim 10$ gravitational radii, would have a size
$\sim 1$~AU.  So in this picture, there would have to be order unity
column density fluctuations in interstellar material on very small length
scales.  In addition, if the change in column
density is due solely to such inhomogeneities, why does it change in the
appropriate sense relative to the continuum flux?
\end{answer}
\begin{question}{Richard Mushotzky}
Does lack of response of CIV, {\it etc.} strongly constrain the model?
\end{question}
\begin{answer}{Julian Krolik}
It's not a very strong constraint for several reasons.  First, the CIV 1549
doublet, and many of the other strong lines, are clearly saturated, so
that their equivalent widths are very insensitive to column density.
Some of the other lines, such as SiIV~1400 are unsaturated (as judged
by their multiplet ratios), and do vary somewhat.  The SiIV line variations
actually resemble the Lyman edge variations, but with somewhat smaller
fractional amplitude.  Second, the fact
that we see a very wide range in ionization states, yet only modest Lyman
edge optical depth, demonstrates that the ionization parameter (the ratio
of ionizing intensity to gas density or pressure) varies considerably
along the line of sight.  So the highly-ionized gas containing the CIV
(or SiIV or $\ldots$) atoms may not be anywhere near the neutral H atoms.
Finally, if
most of the optical depth in the absorption lines is accumulated in regions
where that ionization stage is the dominant one, changes in ionizing flux
don't change the ionization balance very much.  H changes more because
in most photoionized conditions $\Phi \ll 1$.
\end{answer}

\end{document}